\documentstyle[12pt]{article}
\begin{document}
\thispagestyle{empty}
\begin{flushright}
JINR preprint E2-94-39  \\
hep-th/9402067
\end{flushright}
\begin{center}
\Large{ COADDITIVE DIFFERENTIAL COMPLEXES \\
ON QUANTUM GROUPS AND QUANTUM SPACES}
\end{center}

\vspace{.5cm}

\begin{center}
\large{A.A.VLADIMIROV}${}^{\,*}{}^{\,\diamond}$
\end{center}
\begin{center}
\large{Bogolubov Laboratory of Theoretical Physics, \\
Joint Institute for Nuclear Research, \\
Dubna, Moscow region 141980, Russia}
\end{center}

\vspace{1cm}

\begin{center}
ABSTRACT
\end{center}

A regular way to define an additive coproduct (or {\em coaddition})
on the $q$-deformed differential complexes is proposed for quantum
groups and quantum spaces related to the Hecke-type $R$-matrices.
Several examples of braided coadditive differential bialgebras (Hopf
algebras) are presented.

\vspace{6cm}

${}^*$ E-mail: alvladim@thsun1.jinr.dubna.su

\vspace{.3cm}

${}^\diamond$ Work supported in part by the Russian Foundation of
Fundamental Research (grant No. 93-02-3827).

\pagebreak

{\bf 1.} Recently, an additive version of coproduct (or rather
{\em coaddition}) has been observed in various quantum ($q$-deformed)
algebras~\cite{Maj-add,Mey,Maj-coa}. While in the ordinary Lie
algebras this additional algebraic structure is quite natural and
almost trivial, in a $q$-deformed situation it requires
nontrivial braiding rules~\cite{Maj-kit}, thus making the corresponding
quantum algebras the {\em braided coadditive bialgebras} (actually,
Hopf algebras).

A related and very interesting question is a possible bialgebra
structure of differential complexes, i.e., a concept of {\em
differential bialgebras}~\cite{Mal,Man}. Brzezinski~\cite{Br} has shown
that the existence of a bialgebra of this type means the {\em
bicovariance} of the corresponding differential
calculus~\cite{Wo,Ju,AC}.

Therefore, ones interest in the braided coaddition in
differential complexes could be at least threefold:

-- it is interesting by itself, as an additional algebraic structure;

-- it can provide us with a purely Hopf-algebraic criterion for
selecting $q$-deformed differential calculi;

-- it might play a role of a ``shift'' in the physical interpretation
of the corresponding quantum space.

In~\cite{IV}, among other examples, several coadditive differential
bialgebras have been obtained. The aim of the present
paper is to give a systematic approach to this problem for quantum
algebras generated by the $R$-matrices of the Hecke type (for instance,
the $GL_q(N)$ ones~\cite{FRT}). Proceeding in this way, we recover the
results of~\cite{IV}, describe a regular (and very simple) method
to prove consistency (associativity) of the relevant braiding
relations, and find a braided coadditive differential Hopf-algebra
structure on the corresponding quantum group.

This paper has developed from my attempts to interpret
eqs.(\ref{48}),(\ref{49}) (see below) found by A.Isaev~\cite{IV}. I
appreciate this contribution of his to the present work.

\vspace{.3cm}

{\bf 2.} Principal ideas of this paper can be best explained by using
the well accustomed quantum hyperplane
\begin{equation}
R_{12}\,x_1\,x_2=q\,x_2\,x_1  \label{1}
\end{equation}
as an example. We adopt the following notation~\cite{IP,IV}:
\begin{equation}
P_{12}R_{12}\equiv\hat{R}_{12}\equiv R\,, \ \ \ \hat{R}_{23}\equiv R'\,,
\ \ \ R^{-1}\equiv\overline{R}\,, \ \ \ q^{-1}\equiv \bar{q}\,,
\label{2} \end{equation} and also, for any $a$, \begin{equation}
a_1\equiv a\,, \ \ \ a_2\equiv a'\,,\ \ \ a_3\equiv a''\,,\ \ \
a\otimes 1\equiv a\,,\ \ \ 1\otimes a\equiv \tilde{a}\,. \label{3}
\end{equation} For instance, the Yang-Baxter equation and the Hecke
condition for the $R$-matrix look now, respectively, \begin{equation}
R\,R'\,R=R'\,R\,R' \label{4}
\end{equation}
and
\begin{equation}
R-\overline{R}=q-\bar{q}\equiv\lambda  \ \ \ \ {\rm or} \ \ \ \
R^2=1+\lambda R\,. \label{5}
\end{equation}
Our aim is to suppress explicit numerical indices (numbers of the
corresponding auxiliary spaces) in formulae like (\ref{1}) in order not
to mix them with others that we shall need very soon.

Really, the whole differential complex~\cite{WZ} on the quantum
hyperplane (\ref{1}) is defined by
\begin{equation} \label{6}
\left\{ \begin{array}{l}
R\,x\,x'=q\,x\,x'\,, \\
R\,dx\,x'=\bar{q}\,x\,dx'\,, \\
R\,dx\,dx'=-\bar{q}\,dx\,dx'\,.
\end{array} \right.
\end{equation}
Adding formally to this set of equations an extra one,
\begin{equation}
dx\,x'=q\,\overline{R}\,x\,dx'-\lambda \,q\,dx\,x'\,, \label{7}
\end{equation}
which trivially follows from the second line in (\ref{6}), one can
recast (\ref{6}),(\ref{7}) into the matrix form
\begin{equation}
\chi _2\,\chi _1'=Y_{12}\,\chi _1\,\chi _2'\,, \label{8}
\end{equation}
where
\begin{equation}
\chi =\left( \begin{array}{c} x \\ dx \end{array} \right)\,, \ \ \
Y_{12}=q \left( \begin{array}{cccc} \overline{R}&\cdot&\cdot&\cdot \\
\cdot&\overline{R}&-\lambda &\cdot \\ \cdot&\cdot&R&\cdot \\ \cdot&
\cdot&\cdot&-R \end{array} \right)\,, \label{9}
\end{equation}
dots are zeros, and the meaning of numerical indices in (\ref{8}) is,
of course, not the same as in (\ref{1}). It should be noted that the
explicit form (\ref{9}) chosen here for $Y_{12}$ is by no means unique.

Now we are to employ the matrix representation (\ref{8}) for
demonstrating that the differential complex (\ref{6}) admits coaddition
of the form \begin{equation} \Delta(x)=x\otimes 1+1\otimes x\equiv
x+\tilde{x}\,, \ \ \ \ \Delta(dx)=dx+d\tilde{x}\,, \label{10}
\end{equation}
or, in short notation,
\begin{equation}
\Delta(\chi )=\chi +\tilde{\chi}\,. \label{11}
\end{equation}
 From earlier papers on the subject~\cite{Maj-add,IV}, we learn that
this can be only possible when a nontrivial braiding map $\Psi:
\tilde{\Omega }\otimes \Omega\rightarrow \Omega\otimes \tilde{\Omega}$
is used to commute elements with and without a tilde from two
independent copies of our differential complex $\Omega$. Explicitly,
\begin{equation}
(1\otimes a)\,(b\otimes 1)\equiv \tilde{a}\,b=\Psi(a\otimes b)\,.
\label{12}
\end{equation}
In the case (\ref{8}), a natural Ansatz for the braiding is
\begin{equation}
\tilde{\chi }_2\,\chi _1'=Z_{12}\,\chi _1\,\tilde{\chi }_2'\,,
\label{13} \end{equation}
where $Z$ is a $4\!\times\!4$-matrix whose elements may themselves
depend on $R$.

The first restriction on $Z$ is caused by the graded nature of the
differential complex (\ref{6}). This leads to
\begin{equation}
Z_{12}=\left( \begin{array}{cccc} \alpha &\cdot &\cdot&\cdot \\
\cdot&\gamma &\delta &\cdot \\ \cdot&\mu &\beta &\cdot \\
\cdot&\cdot&\cdot&\nu \end{array} \right)\,. \label{14}
\end{equation}

Further, the result of external differentiation of (\ref{13}) must be
consistent with (\ref{13}) itself. Taking into account $d^2=0$ and the
graded Leibnitz rule, we come to
\begin{equation}
\alpha =\beta +\delta \,, \ \ \ \gamma =\delta -\nu \,, \ \ \
\mu =\beta +\nu \,. \label{15}
\end{equation}
The next step is to ensure the key property of $\Delta$, i.e.
\begin{equation}
\Delta(\chi _2)\,\Delta(\chi _1')=Y_{12}\,\Delta(\chi _1)\,\Delta(\chi
_2')\,. \label{16}
\end{equation}
This boils down to verification of
\begin{equation}
\tilde{\chi }_2\,\chi _1'+\chi _2\,\tilde{\chi }_1'=Y_{12}\,
\tilde{\chi }_1\,\chi _2'+Y_{12}\,\chi _1\,\tilde{\chi }_2'\,,
\label{17} \end{equation}
which, with the help of (\ref{13}), transforms to
\begin{equation}
\ [Y_{12}\,Z_{21}+(Y_{12}-Z_{12})P_{12}-{\bf 1}]\,\chi _2\,
\tilde{\chi }_1'=0\,. \label{18}
\end{equation}
We have to put the expression in square brackets to zero. This results
in the following new constraints:
\begin{equation}
\beta =(\delta +1)\,q\,R\,, \ \ \ \ (\nu +1)(R+\bar{q})=0\,. \label{19}
\end{equation}

At last, we must guarantee that our braiding (\ref{13}) obeys so-called
hexagon identities~\cite{Maj-rev} or, equivalently, that our
commutation rules for elements with and without a tilde are
associative. To do this, we perform a reordering
\begin{equation}
\tilde{\chi }_3\,\chi _2'\,\chi _1''\rightarrow \chi _1\,\chi _2'\,
\tilde{\chi }_3'' \label{20}
\end{equation}
in two different ways, using (\ref{8}), (\ref{13}) and
\begin{equation}
\chi _2'\,\chi _1''=Y'_{12}\,\chi _1'\,\chi _2''\,, \ \ \ \ \
\tilde{\chi }_2'\,\chi _1''=Z'_{12}\,\chi _1'\,\tilde{\chi }_2''\,,
\label{21} \end{equation}
where $Y'$ and $Z'$ mean that a substitution $R\rightarrow R'$ in the
corresponding elements of $Y$ and $Z$ has to be carried out. Following
this strategy, we finally obtain
\begin{equation}
Y'_{12}\,Z_{13}\,Z'_{23}=Z_{23}\,Z'_{13}\,Y_{12}\,. \label{22}
\end{equation}
(A similar relation for $Y$,
\begin{equation}
Y'_{12}\,Y_{13}\,Y'_{23}=Y_{23}\,Y'_{13}\,Y_{12}\,, \label{23}
\end{equation}
which expresses the associativity of the original algebra (\ref{6}), is
of course readily verified).

Rewriting the matrix relations (\ref{22}) in the component form, we
immediately encounter
\begin{equation}
R'\,(\beta +\nu )\,\delta '=\delta\,(\beta '+\nu ')\,\overline{R}=0\,.
\label{24} \end{equation}
The only way out is to nullify $\delta $ or $\beta +\nu $. Let us first
consider the latter possibility. Then, due to (\ref{19}),
\begin{equation}
\nu =-\beta \,, \ \ \ \ (\beta -1)(R+\bar{q})=0\,, \ \ \ \
\beta +\delta =\bar{q}\,\overline{R}\,, \label{25}
\end{equation}
and the matrix $Z_{12}$ becomes
\begin{equation}
Z_{12}=\left( \begin{array}{cccc} \bar{q}\overline{R}&\cdot
&\cdot&\cdot \\ \cdot&\bar{q}\overline{R}&
\bar{q}\overline{R}-\beta &\cdot \\ \cdot&\cdot &\beta &\cdot \\
\cdot&\cdot&\cdot&-\beta  \end{array} \right)\,.  \label{26}
\end{equation}
The remaining relations hidden in (\ref{22}) yield
\begin{equation}
\beta \,\overline{R}'\,\overline{R}=\overline{R}'\,\overline{R}\,
\beta '\,, \ \ \ \ \beta \,\beta '\,R=R'\,\beta \,\beta '\,. \label{27}
\end{equation}
The first of these is identically true whereas the second, together
with (\ref{25}), produces two solutions for $\beta $,
\begin{equation}
\beta =\bar{q}R \ \ \ {\rm or} \ \ \beta =q\overline{R}\,, \label{28}
\end{equation}
and, consequently, two possibilities for $Z$,
\begin{equation}
Z_{12}^{(1)}=\bar{q} \left( \begin{array}{cccc}
\overline{R}&\cdot &\cdot&\cdot \\ \cdot&\overline{R}&
-\lambda &\cdot \\ \cdot&\cdot &R &\cdot \\
\cdot&\cdot&\cdot&-R  \end{array} \right), \ \ \ \
Z_{12}^{(2)}=\overline{R} \left( \begin{array}{cccc}
\bar{q}&\cdot &\cdot&\cdot \\ \cdot&\bar{q}&
-\lambda &\cdot \\ \cdot&\cdot &q &\cdot \\
\cdot&\cdot&\cdot&-q  \end{array} \right)\,.  \label{29}
\end{equation}
In the explicit form this reads:
\begin{equation}
\label{30}
\left\{ \begin{array}{l}
\tilde{x}\,x'=\bar{q}\,\overline{R}\,x\,\tilde{x}'\,,  \\
d\tilde{x}\,x'=\bar{q}\,\overline{R}\,x\,d\tilde{x}'
-\lambda \,\bar{q}\,dx\,\tilde{x}'\,,  \\
\tilde{x}\,dx'=\bar{q}\,R\,dx\,\tilde{x}'\,,  \\
d\tilde{x}\,dx'=-\bar{q}\,R\,dx\,d\tilde{x}'\,;
\end{array} \right.
\end{equation}

\vspace{.3cm}

\begin{equation}
\label{31}
\left\{ \begin{array}{l}
\tilde{x}\,x'=\bar{q}\,\overline{R}\,x\,\tilde{x}'\,,  \\
d\tilde{x}\,x'=\bar{q}\,\overline{R}\,x\,d\tilde{x}'
-\lambda \,\overline{R}\,dx\,\tilde{x}'\,,  \\
\tilde{x}\,dx'=q\,\overline{R}\,dx\,\tilde{x}'\,,  \\
d\tilde{x}\,dx'=-q\,\overline{R}\,dx\,d\tilde{x}'\,.
\end{array} \right.
\end{equation}

The other solution of (\ref{24}), $\delta =0$, produces matrices
$\overline{Z}_{21}^{(1)}$ and $\overline{Z}_{21}^{(2)}$ instead of
(\ref{29}). This evidently corresponds to changing the position of a
tilde ($\tilde{\chi }\leftrightarrow \chi , \tilde{x}\leftrightarrow
x$) in (\ref{13}), (\ref{30}) and (\ref{31}), i.e., to the inverse
braiding transformation $\Psi^{-1}$. We thus recover the results
of~\cite{IV} and, moreover, prove that they exhaust all the allowed
braiding relations within the homogeneous Ansatz (\ref{13}). It should
be also stressed that the representations like (\ref{8}) and (\ref{13})
are extremely convenient for proving associativity (resp. consistency)
of appropriate multiplication or braiding relations.

\vspace{.3cm}

{\bf 3}. Now we proceed to the case of the braided matrix algebra
$BM_q(N)$~\cite{KS,Maj-ex} with the generators
$\{1,u^i_j\}$, forming the $N\!\times\!N$-matrix $u$, and relations
\begin{equation}
R_{21}\,u_2\,R_{12}\,u_1=u_1\,R_{21}\,u_2\,R_{12}\,.  \label{32}
\end{equation}
The corresponding differential complex is described in~\cite{OSWZ,AKR}.
In our conventions (note $u_1\equiv u$) it reads
\begin{equation} \label{33}
\left\{ \begin{array}{l}
R\,u\,R\,u=u\,R\,u\,R\,, \\
R\,u\,R\,du=du\,R\,u\,\overline{R}\,, \\
R\,du\,R\,du=-du\,R\,du\,\overline{R}
\end{array} \right.
\end{equation}
(unlike (\ref{6}), there are no primes in these equations). The
appropriate coaddition is also known (see~\cite{Mey} for the $BM_q(N)$
itself and~\cite{IV} for (\ref{33}) as a whole). Here we wish to
reproduce the results of~\cite{IV} through the matrix formalism
developed in the previous section.

Let us rewrite(\ref{33}) in the form
\begin{equation}
\varphi _2\,R\,\varphi _1=V_{12}\,\varphi _1\,R\,\varphi _2\,R\,,
\label{34} \end{equation}
where
\begin{equation} \label{35}
\varphi =\left( \begin{array}{c} u \\ du \end{array} \right), \ \ \ \
V_{12}= \left( \begin{array}{cccc}
\overline{R}&\cdot &\cdot&\cdot \\ \cdot&R&
\cdot &\cdot \\ \cdot&-\lambda  &\overline{R} &\cdot \\
\cdot&\cdot&\cdot&-R  \end{array} \right),
\end{equation}
and try to introduce the braiding relations
\begin{equation}
\tilde{\varphi} _2\,R\,\varphi _1=W_{12}\,\varphi _1\,R\,
\tilde{\varphi _2}\,R\,, \label{36}
\end{equation}
which make
\begin{equation}
\Delta(\varphi )=\varphi +\tilde{\varphi } \label{37}
\end{equation}
a consistent coproduct. From (\ref{34}) and (\ref{36}) we deduce
\begin{equation}
W_{12}\,\varphi _1\,R\,\tilde{\varphi }_2\,R+\varphi _2\,R\,\tilde
{\varphi }_1=V_{12}\,W_{21}\,\varphi _2\,R\,\tilde{\varphi }_1\,R^2
+V_{12}\,\varphi _1\,R\,\tilde{\varphi }_2\,R\,. \label{38}
\end{equation}
With the help of the Hecke condition (\ref{5}) we get
\begin{equation}
(V_{12}\,W_{21}-{\bf 1})\,\varphi _2\,R\,\tilde{\varphi }_1+[\lambda
V_{12}\,W_{21}+(V_{12}-W_{12})P_{12}]\,\varphi _2\,R\,\tilde{\varphi
}_1\,R=0\,.  \label{39}
\end{equation}
A solution is
\begin{equation}
W_{12}=\overline{V}_{21}= \left( \begin{array}{cccc}
R&\cdot &\cdot&\cdot \\ \cdot&R&
\lambda &\cdot \\ \cdot&\cdot &\overline{R} &\cdot \\
\cdot&\cdot&\cdot&-\overline{R}  \end{array} \right)\,.
\label{40}
\end{equation}

Another possible braiding is
\begin{equation}
\tilde{\varphi} _2\,R\,\varphi _1=V_{12}\,\varphi _1\,R\,
\tilde{\varphi _2}\,\overline{R}\,, \label{41}
\end{equation}
inspired by the following equivalent version of (\ref{34}):
\begin{equation}
\varphi _2\,R\,\varphi _1=\overline{V}_{21}\,\varphi _1\,R\,\varphi
_2\,\overline{R}\,. \label{42}
\end{equation}
Of course, this corresponds to the inverse braiding map with respect
to (\ref{36}),(\ref{40}).

Another pair of mutually inverse solutions can be obtained if one
represents (\ref{33}) as
\begin{equation}
\eta _2\,R\,\eta _1=R\,\eta _1\,R\,\eta _2\,V_{12}^{T}=
\overline{R}\,\eta _1\,R\,\eta _2\,\overline{V}_{21}^{T}\,, \label{43}
\end{equation}
where $\eta $ is now a row instead of a column:
\begin{equation}
\eta =(u\;,\;du)\,, \ \ \ \ \
V_{12}^T= \left( \begin{array}{cccc}
\overline{R}&\cdot &\cdot&\cdot \\ \cdot&R&
-\lambda &\cdot \\ \cdot&\cdot &\overline{R} &\cdot \\
\cdot&\cdot&\cdot&-R  \end{array} \right)\,. \label{44}
\end{equation}
In this case, both
\begin{equation}
\tilde{\eta}_2\,R\,\eta _1=R\,\eta _1\,R\,\tilde{\eta}_2\,
\overline{V}_{21}^{T} \label{45}
\end{equation}
and
\begin{equation}
\tilde{\eta}_2\,R\,\eta _1=\overline{R}\,\eta _1\,R\,\tilde{\eta}_2\,
V_{12}^{T} \label{46}
\end{equation}
are consistent braiding relations. Associativity of
(\ref{36}),(\ref{41}),(\ref{45}) and (\ref{46}) (i.e. the identities
like $W_{12}\,W'_{13}\,V_{23}=V'_{23}\,W_{13}\,W'_{12}$)
and their compatibility with the Leibnitz rule are easily confirmed.

In the component form, (\ref{36}) and (\ref{45}) look, respectively, as
\begin{equation} \label{48}
\left\{ \begin{array}{l}
\tilde{u}\,R\,u=R\,u\,R\,\tilde{u}\,R\,, \\
d\tilde{u}\,R\,u=R\,u\,R\,d\tilde{u}\,R+\lambda
\,du\,R\,\tilde{u}\,R\,, \\
\tilde{u}\,R\,du=\overline{R}\,du\,R\,\tilde{u}\,R\,, \\
d\tilde{u}\,R\,du=-\overline{R}\,du\,R\,d\tilde{u}\,R\,;
\end{array} \right.
\end{equation}

\vspace{.3cm}

\begin{equation} \label{49}
\left\{ \begin{array}{l}
\tilde{u}\,R\,u=R\,u\,R\,\tilde{u}\,R\,, \\
d\tilde{u}\,R\,u=R\,u\,R\,d\tilde{u}\,R+\lambda \,R
\,du\,R\,\tilde{u}\,, \\
\tilde{u}\,R\,du=R\,du\,R\,\tilde{u}\,\overline{R}\,, \\
d\tilde{u}\,R\,du=-R\,du\,R\,d\tilde{u}\,\overline{R}\,;
\end{array} \right.
\end{equation}
eqs. (\ref{41}) and (\ref{46}) being obtained from these via
$u\leftrightarrow \tilde{u}$. We recover the corresponding results
given in~\cite{IV}.

\vspace{.3cm}

{\bf 4}. Consider at last the familiar matrix quantum group
\begin{equation}
R_{12}\,T_1\,T_2=T_2\,T_1\,R_{12}\,, \label{50}
\end{equation}
which also has a braided coaddition~\cite{Maj-coa}. Its differential
complex is known too~\cite{Su}. In the notation (\ref{2}),(\ref{3}) it
looks like
\begin{equation} \label{51}
\left\{ \begin{array}{l}
R\,T\,T'=T\,T'\,R\,, \\
R\,dT\,T'=T\,dT'\,\overline{R}\,, \\
R\,dT\,dT'=-dT\,dT'\,\overline{R}\,.
\end{array} \right.
\end{equation}
Let us show that the algebra (\ref{51}) as a whole admits a coaddition
\begin{equation}
\Delta(\theta )=\theta +\tilde{\theta }\,, \ \ \ \ \ \theta \equiv
\left( \begin{array}{c}T \\ dT \end{array} \right)\,. \label{52}
\end{equation}
Really, eq.(\ref{51}) is easily rewritten as
\begin{equation}
\theta _2\,\theta _1'=N_{12}\,\theta _1\,\theta _2'\,R\,, \ \ \ \ \
N_{12}= \left( \begin{array}{cccc}
\overline{R}&\cdot &\cdot&\cdot \\ \cdot&\overline{R}&
-\lambda &\cdot \\ \cdot&\cdot &R &\cdot \\
\cdot&\cdot&\cdot&-R  \end{array} \right)\,. \label{53}
\end{equation}
In complete analogy with the preceding section, one finds that
the mutually inverse braiding relations
\begin{equation}
\tilde{\theta}_2\,\theta _1'=\overline{N}_{21}\,\theta _1\,
\tilde{\theta }_2'\,R\,, \label{54}
\end{equation}
\begin{equation}
\tilde{\theta}_2\,\theta _1'=N_{12}\,\theta _1\,
\tilde{\theta }_2'\,\overline{R} \label{55}
\end{equation}
satisfy all the requirements. If, otherwise, eq.(\ref{51}) is recast
into the form
\begin{equation}
\xi _2\,\xi _1'=R\,\xi _1\,\xi _2'\,N_{12}^{T} \label{56}
\end{equation}
with $\xi $ being a row, $\xi =(T\;,\;dT)$, then the following pair
of mutually inverse braidings is produced:
\begin{equation}
\tilde{\xi}_2\,\xi _1'=R\,\xi _1\,\tilde{\xi}_2'\,
\overline{N}_{21}^{T}\,, \label{57}
\end{equation}
\begin{equation}
\tilde{\xi}_2\,\xi _1'=\overline{R}\,\xi _1\,\tilde{\xi}_2'\,
N_{12}^{T}\,. \label{58}
\end{equation}
In the component form:
\begin{equation} \label{59}
\left\{ \begin{array}{l}
\tilde{T}\,T'=R\,T\,\tilde{T}'\,R\,, \\
d\tilde{T}\,T'=\overline{R}\,T\,d\tilde{T}'\,R\,, \\
\tilde{T}\,dT'=R\,dT\,\tilde{T}'\,R+\lambda
\,T\,d\tilde{T}'\,R\,, \\
d\tilde{T}\,dT'=-\overline{R}\,dT\,d\tilde{T}'\,R\,;
\end{array} \right.
\end{equation}

\vspace{.3cm}

\begin{equation} \label{60}
\left\{ \begin{array}{l}
\tilde{T}\,T'=R\,T\,\tilde{T}'\,R\,, \\
d\tilde{T}\,T'=R\,T\,d\tilde{T}'\,\overline{R}\,, \\
\tilde{T}\,dT'=R\,dT\,\tilde{T}'\,R+\lambda \,R
\,T\,d\tilde{T}'\,, \\
d\tilde{T}\,dT'=-R\,dT\,d\tilde{T}'\,\overline{R}\,;
\end{array} \right.
\end{equation}
two other sets are obtained from these by $\tilde{T}
\leftrightarrow T$.

All the above examples lead us to the conclusion that the braided
coaddition appears to be a quite natural algebraic structure for
the differential complexes on the quadratic quantum algebras generated
by the Hecke-type $R$-matrices. The corresponding (braided) counit
obeys $\varepsilon (1)=1$ and equals zero on other generators.
Moreover, a braided antipode is easily introduced:
\begin{equation}
S(1)=1\,, \ \ \ S(a)=-a\,, \ \ \ S(da)=-da \ \ \ \ \ \ (a=x\,,u\,,T).
\label{61} \end{equation} Consequently, all the braided coadditive
differential bialgebras considered in this paper are, in fact, braided
Hopf algebras.

\end{document}